# INSiDER: Incorporation of System and Safety Analysis Models using a Dedicated Reference Model

Marc Zeller, Siemens AG, Corporate Technology

Kai Höfig, Siemens AG, Corporate Technology

## SUMMARY & CONCLUSIONS

In order to enable model-based, iterative design of safety-relevant systems, an efficient incorporation of safety and system engineering is a pressing need. Our approach interconnects system design and safety analysis models efficiently using a dedicated reference model. Since all information are available in a structured way, traceability between the model elements and consistency checks enable automated synchronization to guarantee that information within both kind of models are consistent during the development life-cycle.

## 1  INTRODUCTION

Nowadays, embedded systems are omnipresent in the daily life, e.g. in industrial automation, medical devices or transportation systems. These systems often implement safety-relevant functionalities. A failure within these systems may lead to catastrophic accidents. Therefore, safety must be considered during the whole development process of a safety-relevant embedded system [1]. However, as practical experiences in industrial projects show, safety analyses are often performed only once and very late in the development cycle because they are very time-consuming.

Traditionally, safety analyses are performed by safety engineers, while the system is designed by system engineers. Thus, separate techniques and tools are used in industry for the design of the system and the analysis of the system in terms of safety. In general, system engineers are not familiar with the methodologies, notions and tools related to safety and vice versa. This is because system and safety engineering are two very different disciplines. However, this gap poses an additional issue during the development of safety-relevant systems and increases the communication as well as the synchronization overhead between the different kinds of engineers.

With the increasing system complexity of today's embedded systems, also the number of safety-relevant functions grows continuously. Due to this, the safety analysis of modern embedded systems has become very complex and time-consuming. To cope with the system complexity in the system development, model-based approaches provide promising solutions [2]. In order to take full advantage of this potential in the development of safety-relevant embedded systems, model-based techniques should be applied for both, the system and the safety engineering [3,4]. Moreover, an iterative and incremental design process should be applied to break with delaying "*try and error*" and "*clone and own*" project cultures in industrial practice and to make the right development decisions in early design phases.

In order to enable model-based, iterative design of safety-relevant embedded systems, the incorporation of safety and system engineering is a pressing need. But to interconnect system and safety analysis models efficiently the following challenges must be solved:

- An automated mapping between the elements of the system design and the safety analysis model is required in order to enable seamless traceability.
- The information within both kinds of models must be kept consistent during the complete development process. For instance, if a certain system element is deleted or renamed, the safety analysis model must be adjusted accordingly. This synchronization should be performed automatically to guarantee that the safety as well as the system engineer always works on consistent data.
- Pre-existing methodologies for both system and safety modeling should be used when incorporating system design and safety analysis. Thus, the system as well as the safety engineer can continue to work with well known techniques and tools, they are already familiar with.

To cope with these challenges, our approach interconnects system design and safety analysis models by using a dedicated reference model. This concept aims to incorporate system and safety engineering during the model-based development process using existing, well-established system design and safety analysis techniques (such as Fault Tree analysis, Markov chains, FMEAs, etc.). Thus, available information from the system model can be used as input for the safety analysis model. Since all information are available in a structured way, the resulting traceability between the model elements enables the automated synchronization to guarantee that information within both kind of models are consistent w.r.t. correctness and completeness.

The reminder of this paper is organized as follows: In Section 2 we summarize related work in this area. Afterwards, our approach to interconnect and synchronize system and safety analysis models with a dedicated reference model is presented and illustrated using a running example. Finally, the paper is concluded in Section 5.

## 2 RELATED WORK

A straight-forward approach to connect system and safety models is to use *model-to-model transformations*. A model transformation has one or several models as input and specifies rules how to produce a specific output model. This method can be adopted to any kind of system or safety model. But since a model-to-model transformation is executed manually, the consistency between system and safety model cannot be guaranteed. Moreover, a specific model-to-model transformation for each kind of system model in combination with each kind of safety analysis is needed.

Various approaches incorporate the safety analysis model into the system model by extending it with specific safety-related information. For instance, a UML-based [5,6], SysML-based [7,8] or EAST-ADL-based [9] model of the system is annotated with safety information by incorporation them into the existing modeling artifacts. Other approaches as presented in [10,11,12] extend the system model by specific modeling artifacts to represent safety information. The safety analysis model may then be derived (semi-)automatically from such an extended system model in order to improve consistency [13]. However, the system model must be extended in order to enable the annotation with safety information. Therefore, specific tooling is needed to work with the extended system model. Moreover, the annotation must be performed manually by the safety engineer, which may not familiar with the common approaches for system modeling. In contrast to this tight coupling of system and safety models, our approach enables loose coupling and supports any kind of system and safety analysis model. Thus, the safety as well as the system engineering can work with models they are familiar with.

Other approaches in the area of model-based safety analysis, such as HiP-HOPS [14] or FPTN [15], are based on a dedicated model for safety analysis. To enable traceability w.r.t. the elements within the system design, the safety analysis model reflects the structure of the system by the notion of components, their relationship, and their hierarchical organization. However, these are specific methodologies which require dedicated tools for safety analysis.

## 3 THE INSIDER APPROACH

In this section, we present our approach to interconnect and synchronize system design and safety analysis models. In the following, this method is described formally and illustrated using an example system (see Figure 1).

### 3.1 Interconnection

In order to enable a model-based and iterative development process for safety-relevant embedded systems, a dedicated reference model is used to link system and safety analysis models. Thus, it stores all information from the system design as well as all safety-related information in a structured way by providing references to these models. The so-called *System Safety Analysis Model ($S^2AM$)* is a common super-set of the *System Model (SM)* (e.g. defined by a generic approach for model-based system engineering such as SysML [16] or a domain-specific one such as EAST-ADL [17] in the automotive domain) as well as the *Safety Analysis Model (SAM)* (e.g. FEMA, Fault Tree or Markov chain):

$$S^2AM \supset \{SM \cup SAM \mid SM \cap SAM = \{\}\} \quad (1)$$

with

$$SM = (C, P, CON) \quad (2)$$

and

$$SAM = (E, FM, \beta) \quad (3)$$

The SM is defined as a tuple consisting of a set of components

$$C = \{c_1, ..., c_n\} \quad (4)$$

and a set of ports

$$P = \{P^{in} \cup P^{out} \mid P^{in} \cap P^{out} = \{\}\} \quad (5)$$

where $P^{in} = \{p_1^{in}, ..., p_r^{in}\}$ is the set of input ports and $P^{out} = \{p_1^{out}, ..., p_s^{out}\}$ is the set of output ports. Each port is allocated to a specific component:

$$\rho : P \to C \quad (6)$$

CON is a set of directed communication links between the components of the systems, respectively their ports:

$$CON \subseteq P^{out} \times P^{in} \quad (7)$$

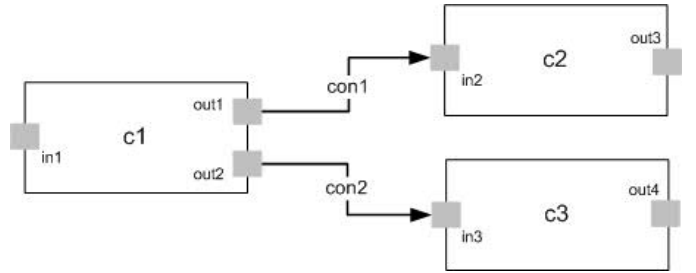

*Figure 1 - Exemplary system model $SM_{Ex}$*

An example system is shown in Figure 1 with

$$SM_{Ex} = \begin{pmatrix} \{c1, c2, c3\}, \\ \{in1, in2, in3, out1, out2, out3, out4\}, \\ \{con1, con2\} \end{pmatrix} \quad (8)$$

with

$$\begin{array}{c} \rho(in1) = c1, \rho(in2) = c2, \rho(in3) = c3, \\ \rho(out1) = c1, \rho(out2) = c1, \\ \rho(out3) = c2, \rho(out4) = c3 \end{array} \quad (9)$$

and

$$con1 = (out1, in2), con1 = (out2, in3) \qquad (10)$$

The SAM is a tuple with a set of (basic) events (also called causes)

$$E = \{e_1, ..., e_r\} \qquad (11)$$

and a set of failure ports

$$F = \{F^{in} \cup F^{out} \mid F^{in} \cap F^{out} = \{\}\} \qquad (12)$$

where $F^{in} = \{f_1^{in}, ..., f_v^{in}\}$ is the set of failure inports and $F^{out} = \{f_1^{out}, ..., f_w^{out}\}$ is the set of failure outports. Moreover, the function

$$\beta : E \cup F^{in} \to F^{out} \qquad (13)$$

describes the relationship between failure inports or (basic) events and failure outports using the Boolean operators $\land, \lor, \neg$.

Apart from describing a complete system by one single SAM, it is also possible to define a SAM for each of the system components:

$$c_i \in C: \quad SAM_{c_i} = (E_{c_i}, F_{c_i}, \beta_{c_i}) \qquad (14)$$

with

$$SAM = \left(\left\{\bigcup_{i=1}^{n} SAM_{c_i} \mid \bigcap_{i=1}^{n} SAM_{c_i} = \{\}\right\}, CON'\right) \qquad (15)$$

CON' is a set of directed communication links describing the failure propagation from the SAM of one component to another:

$$CON' \subseteq F_{c_i}^{out} \times F_{c_j}^{in} \quad \text{with } c_i \neq c_j \in C \qquad (16)$$

Thereby, it is possible to describe the failure propagation within the system in a structured way which is consistent to component-based system design, e.g. by using *Interface Focused-FMEA (IF-FMEA)* [18] or *Component Fault Trees (CFTs)* [19].

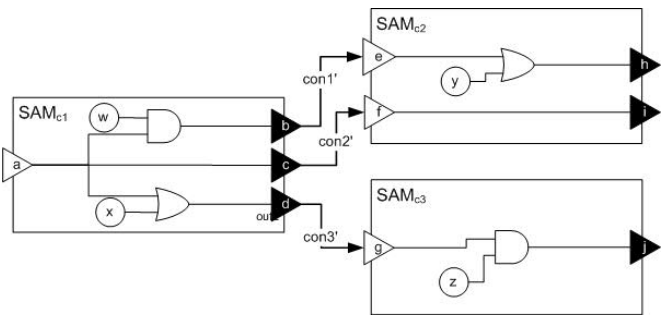

*Figure 2 - Exemplary safety analysis model $SAM_{Ex}$ in form of a CFT*

An exemplary CFT (equivalent to the classical Fault Tree [19]) for the analysis of $SM_{Ex}$ is shown in Figure 2 with

$$SAM_{Ex} = (\{SAM_{c1}, SAM_{c2}, SAM_{c3}\}, \\ \{con1', con2', con3'\}) \qquad (17)$$

where

$$SAM_{c1} = (\{w, x\}, \{a, b, c, d\}, \{a \land w = b, a \lor x = d, a = c\})$$
$$SAM_{c2} = (\{y\}, \{e, f, h, i\}, \{e \lor y = h, f = i\}) \qquad (18)$$
$$SAM_{c3} = (\{z\}, \{g, j\}, \{g \land z = j\})$$

and

$$con1' = (b, e), con2' = (c, f), con3' = (d, g) \qquad (19)$$

In order to enable the seamless information flow between the system and the safety analysis models, the modeling elements can be mapped automatically within the $S^2AM$ by using unique identifiers (e.g. a unique name of the element). For instance, if a component has a unique name within the system model, it can also be identified within the safety analysis model by this name and vice versa. Thus, the relationship between the artifacts within the system model and the safety analysis models is completely traceable. However, the modeling environment (i.e. the tools used to create and maintain the models) must ensure that the names of the modeling elements are unique.

### 3.2 Synchronization

By using our approach, it is also possible to check the consistency of the safety analysis model w.r.t. the given system model. Thus, it can be checked, if

- a representation $SAM_c$ exists within the safety analysis model for each component $c$ within the system model (e.g. a CFT element within a Component Fault Tree):

$$\forall c_i \in C: \quad \exists SAM_{c_i} \qquad (20)$$

- all ports $P$ of each component of the system are represented within the safety analysis model by at least one failure port $f$:

$$\forall p_i^{in} \in P^{in} : \exists f_{c_j}^{in} \in SAM_{c_j} \text{ with } c_j = \rho(p_i^{in}) \qquad (21)$$

and

$$\forall p_i^{out} \in P^{out} : \exists f^{out} \in SAM_{c_j} \text{ with } c_j = \rho(p_i^{out}) \qquad (22)$$

Hence, a mapping between ports and failure ports is created with

$$\gamma: \quad P_{c_i} \to F_{c_i} \qquad (23)$$

$$\gamma': \quad F_{c_i} \to P_{c_i} \qquad (24)$$

- all connections between the ports within the system model are represented by connections between failure ports within the safety analysis model:

$$\forall con \in CON \text{ with } (p_x^{out}, p_y^{in}):$$
$$\exists con \in CON' = \left(\beta_{c_i}\left(E_{c_i} \cup F_{c_i}^{in}\right), f_k^{In}\right)$$
$$\text{with } c_i = \rho(p_x^{out}), c_j = \rho(p_y^{in}), c_i \neq c_j \in C \qquad (25)$$
$$\text{and } f_k^{In} \in F_{c_j}^{In}$$

With the help of these consistency checks, it is possible to detect deviation or gaps between the system design and the safety analysis model.

Based on these consistency checks, an algorithm for the automatic synchronization of the system and the safety analysis model can be build. Thereby, the safety analysis model is synchronized with an existing system model. Hence, a new safety analysis model, which is consistent with the system design, can be created easily (and semi-automatically). Moreover, the safety analysis model can be maintained consistent automatically during the development life-cycle with an evolving system design.

For the synchronization of an existing system design model with a safety analysis model the following algorithm (displayed in pseudo code) is used:

```
FOR each component c in the SM
  IF no SAM_c for component c exists in the SAM THEN
    CALL create new SAM_c in the SAM
  END IF
  FOR each port p of component c
    IF port p is not represented in the SAM_c THEN
      CALL create new failure port f = γ(p) the SAM_c
    END IF
  END FOR
END FOR

FOR each component SAM_c in the SAM
  IF no component c for SAM_c exists in the SM THEN
    CALL remove SAM_c from the SAM
  END IF
  FOR each failure port f of SAM_c
    IF failure port f is not represented in component c
    THEN
      CALL remove failure port f from SAM_c
    END IF
  END FOR
END FOR

FOR each connection con between p1 and p2 in the SM
  IF no connection con' between γ(p1) and γ(p2) exists
    in the SAM THEN
      CALL create new con' between failure ports
           f1 = γ(p1) and f2 = γ(p2)
  END IF
END FOR

FOR each connection con' between f1 and f2 in the SAM
  IF no connection con between γ'(f1) and γ'(f2)
    exists in the SM THEN
      CALL remove con' from the SAM
  END IF
END FOR
```

For example, the system safety analysis model for the exemplary system $S^2AM_{Ex} \supset \{SM_{Ex} \cup SAM_{Ex}\}$ enables to perform the following synchronization between the system design model $SM_{Ex}$ and the safety analysis model $SAM_{Ex}$:

$$\begin{aligned}
c1 &\Rightarrow SAM_{c1} \\
c2 &\Rightarrow SAM_{c2} \\
c3 &\Rightarrow SAM_{c3} \\
in1 &\Rightarrow a \\
in2 &\Rightarrow e, f \\
in3 &\Rightarrow g \\
out1 &\Rightarrow b, c \\
out2 &\Rightarrow d \\
out3 &\Rightarrow h, i \\
out4 &\Rightarrow j \\
con1 &\Rightarrow con1', con2' \\
con2 &\Rightarrow con3'
\end{aligned} \quad (26)$$

Figure 3 illustrates the synchronization between the exemplary system design and its safety analysis model.

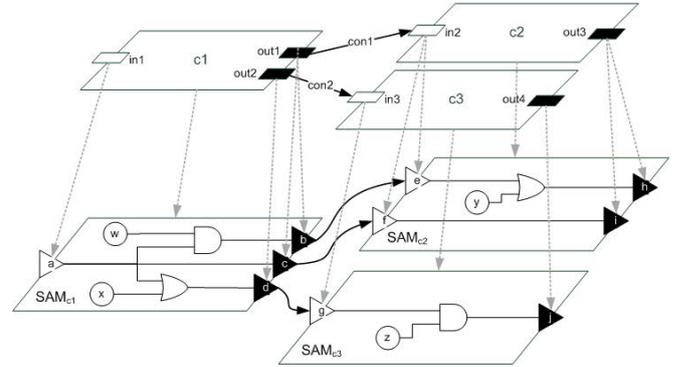

*Figure 3 – Synchronization of the exemplary system design $SA_{Ex}$ and the safety analysis model $SAM_{Ex}$*

### 3.3 Discussion

The $S^2AM$ is composed of a reference to the safety analysis model and a reference to the system model. The use of references enables a loosely coupled linking of the safety analysis models with the system design model without modifying any of the models. An extension of the system model is not needed to store the relevant safety information. Thus, the system engineer and also the safety engineer can work with the notions, methodologies and tools with which they are familiar with. Since the $S^2AM$ is solely working with references, the models can be easily exchanged during the development life-cycle by redefining the references. Also references to additional models can be added in order to integrate additional safety analysis techniques or new system design information without the need to rework existing parts of the model (minimal invasive).

Moreover, the safety analysis model can use information provided in the system model via the references provided by the $S^2AM$. With the information flow enabled by our approach, it is possible to perform safety analyses based on the information from the system design. Hence, the results of the safety analyses are consistent with the system model and can be used as feedback to refine or modify the system design along the development process.

Based on our approach to synchronize system design and safety analysis model, it is possible the keep the information of both kinds of models consistent. Changes of the system design are automatically reflected in the safety analysis model. For instance, if a element in the system design model (component, inport/outport, connection between ports) is added, deleted or renamed, the safety analysis model is adjusted accordingly by adding, deleting or renaming the corresponding elements (safety analysis model, failure inport / failure outport, connection between failure ports). Thus, it is guaranteed that the system engineer as well as the safety engineer works on with the same information, since an automatic mapping between modeling artifacts is possible. However, the synchronization of the elements of the SAM (events and the Boolean formulae describing the intra-component failure propagation) cannot be synchronized automatically, since the information are manually added or modified by the safety engineer. But since these information are clearly assigned to a specific component, it is possible to store them in a repository and reuse them with the component itself.

## 4 CONCLUSON AND FUTURE WORK

Our approach interconnects system design and safety analysis using a dedicated reference model. Thus, available information from the system model can be used as input for the safety analysis model. Since all information are available in a structured way, traceability between the model elements and consistency checks enable an automated synchronization to guarantee that information within both kind of models are consistent. This incorporation of system design and safety analysis models increases the quality of the safety analyses by detection potential errors during the creation and maintenance of the safety analysis models. Furthermore, the time needed for the safety assessment is reduced by enabling semi-automated generation of the analysis models and hence providing modeling support for the system's failure propagation which are kept consistent with the system model. Moreover, safety analyses can be performed iteratively during the whole development life-cycle and provide feedback to the system design, which fosters the incremental, model-based development of safety-relevant embedded systems.

Future work will include the development of methods for the systematic reuse of safety analysis models as well as techniques to fully automate safety analyses at early development stages based on the INSiDER approach.

## ACKNOLEDGEMENT

Parts of the research leading to these results have received funding from the European Union Seventh Framework Programme (FP7 2007-2013) under grant agreement No. 608945 (SafeAdapt).

## REFERENCES

1. International Electrotechnical Commission (IEC), "IEC 61508: Functional safety of electrical/electronic/programmable electronic safety related systems", 1998
2. Holzmann, G. J., "Conquering Complexity", *IEEE Computer*, vol.40, 2007, pp 111-113
3. Joshi, A., Heimdahl, M. P., Miller, S. P., Whalen, M. W., "Model-based safety analysis", *NASA Final Report*, http://shemesh.larc.nasa.gov/fm/papers/Model-BasedSafetyAnalysis.pdf, 2006
4. Schultz, M., Meyer, L., Langer, B., Fricke, H., "Model-based safety assessment as integrated part of system development" *International Workshop on Aircraft System Technologies (AST)*, 2011
5. Le Guennec A., Dion, B., "Bridging UML and safety-critical software development environments", *International Conference on Embedded and Real-Time Software (ERTS)*, 2006
6. Schreiber, S., Schmidberger, T., Fay, A., May, J., Drewes, J., Schnieder, E., "UML-based safety analysis of distributed automation systems", *Proceedings of the IEEE Conference on Emerging Technologies and Factory Automation (ETFA)*, 2007, pp 1069-1075
7. Helle, P., "Automatic SysML-based safety analysis", *Proceedings of the 5th International Workshop on Model Based Architecting and Construction of Embedded Systems (ACES-MB '12)*, 2012, pp 19-24
8. Biggs, G., Sakamoto, T., Kotoku, T., "A profile and tool for modelling safety information with design information in SysML", *Software & Systems Modeling*, 2014, pp 1-32
9. Chen, D., Johansson, R., Lönn, H., Blom, H., Walker, M., Papadopoulos, Y.,Torchiaro, S., Tagliabo, F., Sandberg, A., "Integrated safety and architecture modeling for automotive embedded systems", *e & i Elektrotechnik und Informationstechnik*, vol. 128, 2011, pp 196-202
10. Cancila, D., Terrier, F., Belmonte, F., Dubois, H., Espinoza, H., Gérard, S., Cuccuru, A., "Sophia: A modeling language for model-based safety engineering", *2nd International Workshop on model-based architecting and construction of embedded systems (ACES-MB)*, 2009, pp 11-26
11. Peikenkamp, T., Cavallo, A., Valacca, L., Böde, E., Pretzer, M., Hahn, E.M., "Towards a unified model-based safety assessment", *Computer Safety, Reliability, and Security, Lecture Notes in Computer Science*, vol. 4166, 2006, pp 275-288
12. Zoughbi, G., Briand, L., Labiche, Y., "Modeling safety and airworthiness (RTCA DO-178B) information: conceptual model and UML profile", *Software & Systems*

*Modeling*, vol. 10, 2011, pp 337-367
13. Abdulla, P.A., Deneux, J., Stalmarck, G., Agren, H., Akerlund, O., "Designing safe, reliable systems using SCADE", *Leveraging Applications of Formal Methods, Lecture Notes in Computer Science*, vol. 4313, 2006, pp 115-129
14. Papadopoulos, Y., McDermid, J., "Hierarchically performed hazard origin and propagation studies", *Computer Safety, Reliability and Security*, Springer Berlin Heidelberg, 1999, pp 139-152
15. Fenelon, P., McDermid, J., "An integrated toolset for software safety analysis", *Journal of Systems and Software*, vol. 21, 1993, pp 279-290
16. Object Management Group (OMG), "OMG Systems Modeling Language (OMG SysML)", Version 1.3, http://www.omg.org/spec/SysML/1.3/, 2012
17. EAST-ADL Association, "EAST-ADL Domain Model Specification", Version 2.1.12, http://www.aadl.info/, 2013
18. Papadopoulos, Y., McDermid, J., Sasse, R., Heiner, G., "Analysis and synthesis of the behaviour of complex programmable electronic systems in conditions of failure", *Reliability Engineering & System Safety*, vol. 71, 2001, pp 229-247
19. Kaiser, B., Liggesmeyer, P., Mäckel, O., "A new component concept for fault trees", *Proceedings of the 8th Australian Workshop on Safety Critical Systems and Software (SCS '03)*, 2003, pp 37-46

*BIOGRAPHIES*

Marc Zeller
Siemens AG
Corporate Technology
Otto-Hahn-Ring 6
Munich, 81739, Germany

e-mail: marc.zeller@siemens.com

Marc Zeller works as a research scientist at Siemens AG, Corporate Technology, in Munich since 2014. His research interests are focused on the model-based safety and reliability engineering of complex software-intensive embedded systems. Marc Zeller studied Computer Science at the Karlsruhe Institute of Technology (KIT) and graduated in 2007. He obtained a PhD from the University of Augsburg in 2013 for his work on self-adaptation in networked embedded systems at the Fraunhofer Institute for Embedded Systems and Communication Technologies ESK in Munich.

Kai Höfig
Siemens AG
Corporate Technology
Otto-Hahn-Ring 6
Munich, 81739, Germany

e-mail: marc.zeller@siemens.com

Kai Höfig studied Computer Science at RWTH Aachen, Germany and holds a PhD from the Fraunhofer Institute for Experimental Software Engineering (IESE) where he combined safety-related models and timing analysis models in a probabilistic approach for conditional execution time. He currently leads the Model-based Reliability and Safety Engineering Lab at the Research and Technology Cluster Systems Engineering at Siemens Corporate Technology. There he continues to work with safety-critical systems and supports certification activities in various domains, such as automotive, healthcare, railway, energy and industry automation. His research activities include model-based approaches for reliability, availability, maintainability and safety.